# A Vibrational Approach to Node Centrality and Vulnerability in Complex Networks

Ernesto Estrada[1]

Department of Mathematics and Statistics, Department of Physics, Institute of Complex Systems, University of Strathclyde, Glasgow G1 1XQ, UK.

Naomichi Hatano

Institute of Industrial Science, University of Tokyo, Komaba, Meguro, Tokyo 153-8505, Japan.

[1] Corresponding author. E-mail: ernesto.estrada@strath.ac.uk

**Abstract**

We propose a new measure of vulnerability of a node in a complex network. The measure is based on the analogy in which the nodes of the network are represented by balls and the links are identified with springs. We define the measure as the node displacement, or the amplitude of vibration of each node, under fluctuation due to the thermal bath in which the network is supposed to be submerged. We prove exact relations among the thus defined node displacement, the information centrality and the Kirchhoff index. The relation between the first two suggests that the node displacement has a better resolution of the vulnerability than the information centrality, because the latter is the sum of the local node displacement and the node displacement averaged over the entire network.

## 1. Introduction

The identification of the most central nodes in complex networks has run a long way from the pioneering works of several social scientists interested in quantifying the centrality and prestige of actors in social networks [1-3]. Nowadays, some of these "classical" centrality measures [3], such as degree, betweenness and closeness, play a fundamental role in understanding the structure and properties of complex biological, ecological and infrastructural networks [4-10]. As one of the founding fathers of this field, L.C. Freeman, has stated that the centrality has gone in the "wrong" way: from social sciences to physical science, in contrast with the traditional way, which is from physical science to social ones [11]. A concept that is quite related to that of centrality is the concept of vulnerability [12-16]. In the seminal paper of Albert, Jeong and Barabási [17] the analysis of the vulnerability of a network to intentional attacks is carried out by considering the degree of the nodes in the network. Then, in scale-free networks the removal of the highest-degree nodes makes the network collapse into many small isolated chunks. This idea has been extended by others to consider other centrality measures [12-16]. In this context, for instance, it is possible to identify large parts of the network that can be isolated by removing the nodes with the largest betweenness centrality [12].

A different question, however, is the identification of the most vulnerable nodes in a network. In a network we assume that we can attack any node by simply removing it from the graph. We call this removal a primary extinction. However, it is well known that a node can be completely disconnected from the rest of the network by removing other node(s) in the graph [17]. For instance, consider the case of a node $i$ having only one connection, i.e., its degree is one. In this case removing the node to which $i$ is attached will isolate it from the rest of the nodes. We will refer as a secondary extinction to the effect produced on other nodes by the primary extinction of a given one. Then, it is expected that high-degree nodes are less vulnerable than the low-degree ones. This has been the basis of the strategy of attacking networks by removing hubs used by Albert, Jeong and Barabási [17]. Because the hubs are highly connected nodes, their disconnection will leave many low-degree nodes isolated. Observe that the attack strategy is always to remove the least vulnerable nodes by primary attacks as the most vulnerable ones will be disabled as a consequence of secondary extinctions. However, the node degree accounts for only the nearest-neighbours attached to a node. Consequently, the influence of more distant nodes is not considered in the vulnerability analysis based on the node degree. This has given rise to the use of alternative strategies, such as the use of betweenness centrality [12].

The purpose of the present study is to find a more appropriate and useful measure of vulnerability. In a general context, we can think about the node vulnerability as the susceptibility that a node has to any perturbation taking place on the network. In the present paper we use the physical concept of vibrations to account for the node vulnerability. We submerge the network in question in a thermal bath. The thermal fluctuation works as the "perturbations" acting on the network. As a measure of vulnerability, we will use the displacement of a node from its "equilibrium" position due to small "oscillations" in the network.

In Sec. 2, we will define the node displacement and find its expression in terms of the Moore-Penrose generalised inverse of the graph Laplacian matrix. We will argue in Sec. 3 the role played by the temperature in the present theory. Then in Sec. 4, using spectral graph theory we will show that the node displacement is a good measure of the vulnerability of a node. We will prove analytically in Sec. 5 that the so-called information centrality introduced by Stephenson and Zelen [18] is the sum of the local square node displacement and the average square displacement. Thus, we reinterpret this centrality measure here in terms of the vulnerability of a node to the perturbations taking place in the network. We will also prove in Sec. 6 that the Kirchhoff index is proportional to the average square node displacement. Finally in Sec. 7, we will illustrate the applications of the new concepts developed here by studying a social network. We will show that despite the similarities between the information centrality and the node displacements, the node displacement has a better resolution of vulnerability than the information centrality. We demonstrate this by analysing temporal changes of a real-world network.

## 2. Vibrations in complex networks

First we consider the analogy in which the nodes of the complex network are represented by balls and the links are identified with springs with a common spring constant $k$. We would like to consider the vibrational excitation energy from the static position of the network. Let $x_i$ denote the displacement of a node $i$ from its static position. The meaning of these displacements will be evident later. Then the vibrational potential energy of the network can be expressed as

$$V(\vec{x}) = \frac{k}{2}\vec{x}^T \mathbf{L}\vec{x}, \tag{1}$$

where $\mathbf{L} = \mathbf{D} - \mathbf{A}$ is the discrete Laplacian matrix of the network and $\vec{x}$ is the vector whose $i$th entry is the displacement $x_i$. We recall that $\mathbf{D}$ is the diagonal matrix of degrees and $\mathbf{A}$ is the adjacency matrix of the network. Suppose that the network is affected by an external stress. In our physical model this is simulated by immersing the network into a thermal bath of inverse temperature $\beta$. The meaning of the temperature concept in the context of complex networks will be clarified in the next section. Then the probability distribution of the displacement of the nodes is given by the Boltzmann distribution

$$P(\vec{x}) = \frac{e^{-\beta V(\vec{x})}}{Z} = \frac{1}{Z}\exp\left(-\frac{\beta k}{2}\vec{x}^T \mathbf{L}\vec{x}\right), \tag{2}$$

where the normalization factor $Z$ is the partition function of the network

$$Z \equiv \int d\vec{x} \exp\left(-\frac{\beta k}{2}\vec{x}^T \mathbf{L}\vec{x}\right). \tag{3}$$

The mean displacement of a node $i$ can be expressed by

$$\Delta x_i \equiv \sqrt{\langle x_i^{\,2} \rangle} = \sqrt{\int x_i^{\,2} P(\vec{x}) d\vec{x}}\,, \tag{4}$$

where $\langle \cdots \rangle$ denotes the thermal average. We can also define the displacement correlation as

$$\langle x_i x_j \rangle = \int x_i x_j P(\vec{x}) d\vec{x}. \tag{5}$$

We can calculate this quantity once we diagonalise the Laplacian matrix $\mathbf{L}$. Let us denote by $U$ the matrix whose columns are the orthonormal eigenvectors $\vec{\psi}_\mu$ and $\Lambda$ the diagonal matrix of eigenvalues $\lambda_\mu$ of the Laplacian matrix. Note here that the eigenvalues of the Laplacian of a connected network are positive except for one zero-eigenvalue. Then, we write the Laplacian spectrum as $0 = \lambda_1 < \lambda_2 \leq \cdots \leq \lambda_n$. An important observation here is that the *zero eigenvalue does not contribute to the vibrational energy*. This is because the mode $\mu = 1$ is the mode where all the nodes (balls) move coherently in the same direction and thereby the whole network moves in one direction. In other words, this is the motion of the center of mass, not a vibration.

In calculating Eqs. (2) and (3), the integration measure is transformed as

$$d\vec{x} \equiv \prod_{i=1}^{n} dx_i = |\det U| \prod_{i=1} dy_i = d\vec{y} \tag{6}$$

because the determinant of the orthogonal matrix, $\det U$, is either $\pm 1$. Then we have

$$\begin{aligned} Z &= \int d\vec{y} \exp\left(-\frac{\beta k}{2}\vec{y}^T \Lambda \vec{y}\right) \\ &= \prod_{\mu=1}^{n} \int_{-\infty}^{+\infty} dy_\mu \exp\left(-\frac{\beta k}{2}\lambda_\mu y_\mu^{\;2}\right). \end{aligned} \tag{7}$$

Note again that because $\lambda_1 = 0$ the contribution from this eigenvalue obviously diverges. This is because nothing stops the whole network from moving coherently in one direction. When we are interested in the vibrational excitation energy within the network, we should offset the motion of the center of mass and focus on the relative motion of the nodes. We therefore redefine the partition function by removing the first component $\mu = 1$ from the last product. We thereby have

$$\begin{aligned}\tilde{Z} &= \prod_{\mu=2}^{n} \int_{-\infty}^{+\infty} d y_\mu \exp\left(-\frac{\beta k}{2} \lambda_\mu y_\mu^{\,2}\right) \\ &= \prod_{\mu=2}^{n} \sqrt{\frac{2\pi}{\beta k \lambda_\mu}}.\end{aligned} \tag{8}$$

Next we calculate the mean displacement $\Delta x_i$ defined by Eq. (4). We first compute the numerator of the right-hand side of Eq. (4) as follows:

$$\begin{aligned}I_i &= \int d\vec{x} x_i^{\,2} \exp\left(-\frac{\beta k}{2} \vec{x}^T \Lambda \vec{x}\right) \\ &= \int d\vec{y} (U\vec{y})_i^{\,2} \exp\left(-\frac{\beta k}{2} \vec{y}^T \Lambda \vec{y}\right) \\ &= \int d\vec{y} \left(\sum_{\nu=1}^{n} U_{i\nu} y_\nu\right)^2 \exp\left(-\frac{\beta k}{2} \vec{y}^T \Lambda \vec{y}\right) \\ &= \int d\vec{y} \left(\sum_{\nu=1}^{n} \sum_{\gamma=1}^{n} U_{i\nu} U_{i\gamma} y_\nu y_\gamma\right) \prod_{\mu=1}^{n} \exp\left(-\frac{\beta k}{2} \lambda_\mu y_\mu^{\,2}\right).\end{aligned} \tag{9}$$

On the right-hand side, any terms with $\nu \neq \gamma$ will vanish after integration because the integrand is an odd function with respect to $y_\nu$ and $y_\gamma$. The only possibility of a finite result is due to terms with $\nu = \gamma$. We therefore have

$$I_i = \int d\vec{y} \left[ \sum_{\nu=1}^{n} (U_{i\nu} y_\nu)^2 \right] \prod_{\mu=1}^{n} \exp\left( -\frac{\beta k}{2} \lambda_\mu y_\mu^{\ 2} \right)$$

$$= \int_{-\infty}^{+\infty} dy_1 (U_{i1} y_1)^2 \times \prod_{\mu=2}^{n} \int_{-\infty}^{+\infty} dy_\mu \exp\left( -\frac{\beta k}{2} \lambda_\mu y_\mu^{\ 2} \right) + \tag{10}$$

$$\sum_{\nu=2}^{n} \int_{-\infty}^{+\infty} dy_\nu (U_{i\nu} y_\nu)^2 \exp\left( -\frac{\beta k}{2} \lambda_\nu y_\nu^{\ 2} \right) \times \prod_{\substack{\mu=1 \\ \mu\neq\nu}}^{n} \int_{-\infty}^{+\infty} dy_\mu \exp\left( -\frac{\beta k}{2} \lambda_\mu y_\mu^{\ 2} \right),$$

where we separated the contribution from the zero eigenvalue and those from the other ones. Due to the divergence introduced by the zero eigenvalue we proceed the calculation by redefining the quantity $I_i$ with the first node removed:

$$\tilde{I}_i \equiv \sum_{\nu=2}^{n} \int_{-\infty}^{+\infty} dy_\nu (U_{i\nu} y_\nu)^2 \exp\left( -\frac{\beta k}{2} \lambda_\nu y_\nu^{\ 2} \right) \times \prod_{\substack{\mu=2 \\ \mu\neq\nu}}^{n} \int_{-\infty}^{+\infty} dy_\mu \exp\left( -\frac{\beta k}{2} \lambda_\mu y_\mu^{\ 2} \right)$$

$$= \sum_{\nu=2}^{n} \frac{U_{i\nu}^{\ 2}}{2} \sqrt{\frac{8\pi}{(\beta k \lambda_\nu)^3}} \times \prod_{\substack{\mu=2 \\ \mu\neq\nu}}^{n} \sqrt{\frac{2\pi}{\beta k \lambda_\nu}} \tag{11}$$

$$= \tilde{Z} \times \sum_{\nu=2}^{n} \frac{U_{i\nu}^{\ 2}}{\beta k \lambda_\nu}.$$

We therefore arrive at the following expression for the mean displacement of a node, Eq. (4):

$$\Delta x_i \equiv \sqrt{\langle x_i^{\ 2} \rangle} = \sqrt{\frac{\tilde{I}_i}{\tilde{Z}}} = \sqrt{\sum_{\nu=2}^{n} \frac{U_{i\nu}^{\ 2}}{\beta k \lambda_\nu}}. \tag{12}$$

If we designate by $\mathbf{L}^+$ the Moore-Penrose generalised inverse (or the pseudo-inverse) of the graph Laplacian [8], which has been proved to exist for any molecular graph, then it is straightforward to realise that

$$(\Delta x_i)^2 = \frac{1}{\beta k} (\mathbf{L}^+)_{i\,i}. \tag{13}$$

A similar calculation also gives the displacement correlation, Eq. (5), in the form

$$\langle x_i x_j \rangle = \sum_{\nu=2}^{n} \frac{U_{i\nu} U_{j\nu}}{\beta k \lambda_\nu} = \frac{1}{\beta k} (\mathbf{L}^+)_{i\,j}. \tag{14}$$

Returning to a real-world situation, the displacement of a node given by $\Delta x_i$ represents how much the corresponding node is affected by the external stress to which the network is submitted to. A node which displays a large value of the displacement is one which is very much affected by the external conditions, such as economical crisis, social agitation, environmental pressure or physiological conditions. In other words, it is a node with high *vulnerability* to the change in the external conditions. This equivalence between displacement and vulnerability will be analysed in the next section.

## 3. On the concept of temperature in complex networks

Complex networks are continuously exposed to external "stress" which is independent of the organizational architecture of the network. By external we mean here an effect which is independent of the topology of the network. For instance, let us consider the network in which nodes represent corporations and the links represent their business relationships. In this case the external stress can represent the economical situation of the world at the moment in which the network is analysed. In "normal" economical situations we are under a low level of external stress. In situations of economical crisis the level of external stress is elevated. Despite the fact that these stresses are independent of the topology of the network they can have a determinant role in the evolution of the structure of these systems. For instance, in a situation of high external stress like an economical crisis it is plausible that some of the existing links between corporations are lost at the same time as new merging and strategic alliances are created. The situation is also similar for the social ties between actors in a society, which very much depend on the level of "social agitation" existing in such a society in the specific period of time under study.

An analogy exists between the previously described situation and what happens at the molecular level of organization in a given substance. In this case such organization has a strong dependence on the external temperature. For instance, the intermolecular interactions between molecules differ significantly in gas, liquid and solid states of matter. The agents (molecules) forming these systems are exactly the same but their organization changes dramatically with the change of the external temperature. We are going to use this analogy to capture the external stresses influencing the organization of a complex network.

Suppose that the complex network is submerged into a thermal bath at the temperature $T$. The thermal bath represents here an external situation which affects all the links in the network at the same time. It fulfils the requirements of being independent of the topology of the network and of having a direct influence over it. Then, after equilibration all links in the network will be weighted by the parameter $\beta = (k_B T)^{-1}$. The parameter $\beta$ is known as the *inverse temperature* and $k_B$ is the Boltzmann constant.

This means that when the temperature tends to infinity, $\beta \to 0$, all links have weights equal to zero. In other words, the graph has no links. This graph is known as the trivial or empty graph. For $\beta = 1$, we have the simple graph in which every pair of connected nodes have a single link. The concept of temperature in networks can be considered as a specific case of a weighted graph in which all links have the same weights. If we consider the case when the temperature tends to zero, $\beta \to \infty$, all links have infinite weights. Then, we can have an analogy with the different states of the matter, in which the empty graph obtained at very high temperatures $\beta \to 0$ is similar to a gas, formed by free particles with no link among them. In the other extreme $\beta \to \infty$, we have a graph with infinite weights in their links, which in some way has resemblances with a solid. The simple graph may be then analogous to the liquid state. These kinds of networks, obtained at different temperatures are illustrated in Fig. 1.

**Insert Fig. 1 about here.**

## 4. Node displacement as a measure of node vulnerability

The node displacement $\Delta x_i$ depends on the eigenvectors corresponding to all non-trivial eigenvalues of the Laplacian matrix. Consequently, this measure contains information not only on nearest-neighbours but also on the influence of more distant nodes from $i$.

For instance, in the simple graph illustrated in Fig. 2 the values of node displacement reflect the influences of all nodes on a particular one to be analysed. In this example the node displacements are: $\Delta x_1 = \Delta x_2 = \Delta x_3 = \Delta x_7 = \Delta x_8 = \Delta x_9 = 1.116$, $\Delta x_4 = \Delta x_6 = 0.762$ and $\Delta x_5 = 0.68$, for the nodes with labels given in Fig. 2. This means that the nodes of degree one are very much affected by the external conditions, which means that they are highly vulnerable. It is easy to realise that these nodes are completely isolated by removing nodes 4 or 6. In addition the removal of node 5 leaves nodes of degree one in an isolated cluster containing four nodes.

The second most vulnerable nodes according to the displacement are nodes 4 and 6. They are not affected by the removal of nodes of degree 1 but they are left into an isolated cluster of four nodes when node 5 is removed. Finally, node 5 is the least vulnerable one. It is not affected by the removal of nodes of degree one and when nodes 4 or 6 are removed it still remains in the largest component of the graph, which is formed by five nodes.

Obviously, the degree fails in ranking the vulnerability of these nodes as it identifies nodes 4 and 6 as the least vulnerable ones. In addition, betweenness centrality also fails in identifying the least vulnerable node in this graph. The betweenness of nodes 1, 4 and 6 are 0, 18 and 16, respectively, which also identifies node 4 as the least vulnerable one.

**Insert Fig. 2 about here.**

An explanation of why the displacement is a good measurement of the node vulnerability can be given in terms of the Laplacian spectrum. We recall that we have ordered the eigenvalues of the Laplacian as $0 = \lambda_1 < \lambda_2 \leq \cdots \leq \lambda_n$. Let us consider the particular case $\lambda_2 < \lambda_3$. Then because the eigenvalues of the Laplacian appear in the denominator of the node displacement (12), the term $\frac{U_{i2}^2}{\lambda_2}$ has the largest contribution to $\Delta x_i$. The smallest non-trivial eigenvalue $\lambda_2$ and its corresponding eigenvector are known as the algebraic connectivity and the Fiedler vector of the graph, respectively [19]. The Fiedler vector is very well known in spectral graph partitioning [20]. Accordingly, the nodes of the graph are partitioned into two sets $V_1$ and $V_2$ as follows: $V_1 = \{i \mid U_{i2} < 0\}$, $V_2 = \{i \mid U_{i2} \geq 0\}$. That is, a node belongs "strongly" to a partition if it has a large positive or a large negative value of the corresponding entry of the Fiedler vector, whereas a node with $U_{i2}$ close to zero does not belong strongly to either of the two partitions. Consequently, the latter nodes can be considered as "separators" among the partitions. Then, if we remove such separators we in general isolate the two partitions found by the positive and negative components of the Fiedler vector.

For instance, in the previous example, we have $U_{52}=0$, $U_{42}= 0.086$ and $U_{12}= 0.138$. Of course, the equivalent nodes at the other side of the graph have the same values of the Fiedler vector but with negative signs. As a first approach we can consider a partition of the graph in which the nodes 1, 2, 3 and 4 form a cluster and the nodes 6, 7, 8 and 9 form another. Then, it is clear that the node 5 is a separator of the two partitions. This means that removing this separator will isolate the two clusters. In the second approach we can consider that a cluster is formed by nodes 1, 2 and 3, while the other by nodes 7, 8, 9. In this case, the nodes 4, 5 and 6 are separators. As can be seen the node 5 appears as a separator in the two possible partitions and the nodes 4 and 6 in only one of them, which introduces the ranking in vulnerabilities given by the node displacement.

It is worth mentioning here that the algebraic connectivity gives a measure of the robustness of a graph. The larger the algebraic connectivity, the harder it is to separate a graph into isolated components. It is known that $\lambda_2 \leq \kappa(G)$, where $\kappa(G)$ represents the node connectivity, which is the minimum number of nodes that must be removed in order to disconnect the graph. Then, the two main contributions to the displacement guarantee that the smallest value of $\Delta x_i$ is obtained for the nodes contributing more to separate the network into disconnected parts.

The obvious question at this point is why to use $\Delta x_i$ instead of the Fiedler vector as a measure of node vulnerability. First, it has been previously proved that the use of as many eigenvectors as practically possible should be preferred for partitioning purposes. While the Fiedler vector gives a bipartition of the network the other eigenvectors give partitions into larger number of clusters, which helps to identify the "separators" in a more appropriate way. In the second place, the Fiedler vector may be degenerate. That is, the assumption $\lambda_2 < \lambda_3$ is not always true, e.g., in square grids and complete graphs. In this case it has been observed that there are great problems with the convergence of algorithms used for partitioning purposes. Finally, the use of the physically appealing concept of node vibrations permits us also to compare nodes in different networks on the basis of the penalization imposed by the eigenvalues in the expression for $\Delta x_i$ .

A desired property for a measure of node vulnerability in complex networks is that it accounts in some way for the empirical observation that low-degree nodes are more vulnerable than the high-degree ones. In Fig. 3 we illustrate the relation between the displacement and the degree for all connected graphs having 5 nodes. It can be seen that the node displacement correlates with the node degree as $\Delta x_i \sim k_i^{-\alpha}$ with some power $\alpha = 1.15$, which agrees with the intuition that low-degree nodes are more vulnerable than high-degree ones. This kind of power-law relationship between the node displacement and the node degree is observed for all artificial and real-world networks studied so far. However, for nodes with the same degree a large variability in their displacements is observed, which adds an extra value to the use of $\Delta x_i$ instead of the node degrees as a measure of node vulnerability.

**Insert Fig. 3 about here.**

## 5. Node displacement and information centrality

In the present and next sections we will find exact relations between the node displacement and other measures. A measure of centrality of a node in a network was introduced by Stephenson and Zelen [18] and is known as *information centrality* (*IC*). It is based on the information that can be transmitted between any two points in a connected network. If **A** is the adjacency matrix of a network, **D** a diagonal matrix of the degree of each node and **J** a matrix with all its elements equal to one, then *IC* is defined by inverting the matrix $\mathbf{B} \equiv \mathbf{D} - \mathbf{A} + \mathbf{J}$ as $\mathbf{C} = (c_{ij}) = \mathbf{B}$, from which the information matrix is obtained as follows:

$$\mathbf{I}_{ij} = \left(c_{ii} + c_{jj} - 2c_{ij}\right)^{-1} \tag{15}$$

The information centrality of the species *i* is then defined by using the harmonic average:

$$IC(i) = \left[\frac{1}{n}\sum_{j}\frac{1}{\mathbf{I}_{ij}}\right]^{-1} \tag{16}$$

Stephenson and Zelen [18] proposed to define $\mathbf{I}_{ii}$ as infinite for computational purposes, which makes $1/\mathbf{I}_{ii} = 0$.

It is straightforward to realise that $\mathbf{B} = \mathbf{L} + \mathbf{J}$. The following theoretic result guarantees the existence of $\mathbf{C} = \mathbf{B}^{-1}$ for connected networks.

**Lemma 1**: If $G$ is a connected network, then the inverse of $\mathbf{B}$ has the same eigenvectors as the Laplacian with eigenvalues $n, \lambda_2, \lambda_3, \ldots, \lambda_n$. The inverse of $\mathbf{B}$ is given by

$$\mathbf{B}^{-1} = \mathbf{L}^{+} + \frac{1}{n^{2}}\mathbf{J}. \tag{17}$$

where as before $\mathbf{L}^{+}$ is the Moore-Penrose generalised inverse of $\mathbf{L}$.

*Proof*: Let $k \neq 1$. Then,

$$\mathbf{B}U_k = (\mathbf{L} + \mathbf{J})U_k = \mathbf{L}U_k + \mathbf{J}U_k = \lambda_k U_k, \tag{18}$$

because $\mathbf{J}U_k = \vec{0}$. Now, let $k = 1$. Then, $\mathbf{J}U_1 = nU_1$ due to the fact that $U_1 = \frac{1}{\sqrt{n}}\vec{1}$ and because $\lambda_1 = 0$ and $\mathbf{L}U_1 = \vec{0}$, we have

$$\mathbf{B}U_1 = (\mathbf{L} + \mathbf{J})U_1 = \mathbf{L}U_1 + \mathbf{J}U_1 = nU_1, \tag{19}$$

which means that $n$ is the eigenvalue of $\mathbf{B}$ corresponding to $U_1$.

It is known that $\mathbf{L}\mathbf{L}^{+} = \mathbf{L}^{+}\mathbf{L} = \mathbf{I} - \frac{1}{n}\mathbf{J}$, $\mathbf{L}\mathbf{J} = \mathbf{J}\mathbf{L} = 0$, $\mathbf{L}^{+}\mathbf{J} = \mathbf{J}\mathbf{L}^{+} = 0$ and $\mathbf{J}^{2} = n\mathbf{J}$. Then, we have

$$(\mathbf{L} + \mathbf{J})\left(\mathbf{L}^{+} + \frac{1}{n^{2}}\mathbf{J}\right) = \mathbf{I} - \frac{1}{n}\mathbf{J} + \frac{n\mathbf{J}}{n^{2}} = \mathbf{I}, \tag{20}$$

which proves that $\mathbf{L}^{+} + \frac{1}{n^{2}}\mathbf{J}$ is the inverse of $\mathbf{B}$.

Then, we have the following new result concerning the information centrality.

**Theorem 2**: Let $\Delta x_i$ be the node displacement and $\overline{(\Delta x)^2}$ the average displacement over all nodes in the network. Then, the information centrality of node $i$ is given by

$$IC(i) = \frac{1}{\beta k\left[(\Delta x_i)^2 + \overline{(\Delta x)^2}\right]}. \tag{21}$$

*Proof*: Let us write the information centrality as follows

$$IC(i) = \left[c_{ii} + \frac{(T - 2R)}{n}\right]^{-1}, \tag{22}$$

where $T = \sum_{j=1}^{n} c_{jj}$ and $R = \sum_{j=1}^{n} c_{ij}$. Then,

$$c_{ii} = \left(\mathbf{B}^{-1}\right)_{ii} = \left(\mathbf{L}^{+} + \frac{1}{n^{2}}\mathbf{J}\right)_{ii} = \left(\mathbf{L}^{+}\right)_{ii} + \frac{1}{n^{2}} = \beta k\left(\Delta x_i\right)^{2} + \frac{1}{n^{2}}, \tag{23}$$

$$T=\sum_{j=1}^{n} c_j = \mathrm{Tr}\,\mathbf{B}^{-1} = \mathrm{Tr}\left(\mathbf{L}^{+}+\frac{1}{n^2}\mathbf{J}\right) = \mathrm{Tr}\,\mathbf{L}^{+}+\frac{1}{n} = \beta k\sum_{i=1}^{n}(\Delta x_i)^2+\frac{1}{n}, \tag{24}$$

$$R=\sum_{j=1}^{n} c_i = \sum_{j=1}^{n}\left(\mathbf{B}^{-1}\right)_{ij} = \frac{1}{n}, \tag{25}$$

because $\mathrm{Tr}\,\mathbf{J}=n$ and $\sum_{j=1}^{n}\left(\mathbf{L}^{+}\right)_{ij} = \sum_{i=1}^{n}\left(\mathbf{L}^{+}\right)_{ij} = 0$. Then, we have

$$I(i)=\left[\beta k(\Delta x_i)^2+\frac{1}{n^2}+\frac{\left(\beta k\sum_{i=1}^{n}(\Delta x_i)^2+\frac{1}{n}-\frac{2}{n}\right)}{n}\right]^{-1} = \frac{1}{\beta k\left[(\Delta x_i)^2+\overline{(\Delta x)^2}\right]}. \tag{26}$$

## 6. Electrical networks analogy

Let us consider a connected network which has associated with an electrical network in such a way that each link of the network is replaced by a resistor of electrical resistance equal to one Ohm. When the poles of a battery are connected to any pair of nodes $i$ and $j$ in the network the resulting resistance $\Omega_{ij}$ between them is given by the Kirchhoff and Ohm laws. Such resistance is known to be a distance function [21] and called the resistance distance. It was introduced in a seminal paper by Klein and Randić a few years ago [21] and has been intensively studied in mathematical chemistry [21-25]. The standard method of computing the resistance distance for a pair of nodes in a network is by using the Moore-Penrose generalised inverse of the graph Laplacian $\mathbf{L}^{+}$, which is given by [21, 22]

$$\Omega_{ij}=\left(\mathbf{L}^{+}\right)_{ii}+\left(\mathbf{L}^{+}\right)_{jj}-\left(\mathbf{L}^{+}\right)_{ij}-\left(\mathbf{L}^{+}\right)_{ji}, \tag{27}$$

for $i\neq j$. The resistance matrix $\mathbf{\Omega}$ is the matrix whose non-diagonal entries are $\Omega_{ij}$ and $\Omega_{ii}=0$ [21]. It is easy from Eqs. (12) and (14) to see

$$\Omega_{ij}=\beta k\left[(\Delta x_i)^2+(\Delta x_j)^2-\langle x_i x_j\rangle-\langle x_j x_i\rangle\right]=\beta k\left\langle (x_i-x_j)^2\right\rangle. \tag{28}$$

The right-hand side of Eq. (28) should be small when the nodes $i$ and $j$ vibrate in the same direction and should be large when they vibrate in the opposite directions. If we focus particularly on the mode corresponding to $\lambda_2$, on which we discussed in Sec. 4, two nodes in the same partition according to the Fiedler vector vibrate in the same direction and ones in different partitions vibrate in the opposite directions. Then, Eq. (28) dictates that two nodes that are close in terms of the resistance distance tend to be in the same partition and vibrate in the same direction whereas ones that are far tend to be in different partitions and vibrate in the opposite directions, which fits our intuition well.

The resistance matrix $\Omega$ is the basis of a topological index for a network that was introduced first for studying molecular graphs. This index is known as the Kirchhoff index $Kf$ and is defined as $Kf = \sum_{i<j} \Omega_{ij}$ [21-25]. It is known that $Kf$ can be expressed in terms of the Laplacian eigenvalues as follows [22],

$$Kf = n\sum_{j=2}^{n} \frac{1}{\lambda_j} = n \mathrm{Tr}\, \mathbf{L}^{+}, \tag{29}$$

which means that we can express the Kirchhoff index using the node displacements as

$$Kf = n\beta k \sum_{i=i}^{n} (\Delta x_i)^2 = n^2 \beta k \overline{(\Delta x)^2}. \tag{30}$$

This in turn gives the information centrality [18] in the form

$$I(i) = \left[ \left(\mathbf{L}^{+}\right)_{ii} + \frac{Kf}{n^2} \right]^{-1}, \tag{31}$$

where $\left(\mathbf{L}^{+}\right)_{ii} = \beta k (\Delta x_i)^2$ is interpreted here as the contribution of the node to the effective resistance of every link attached to it. In the case of networks having no cycles, i.e., trees, it is known that the Kirchhoff index and the sum of shortest path lengths between all pairs of nodes in the graph coincide [21].

Let $R_i = \sum_j \Omega_{ij}$ be the sum of the resistance distances from node $i$ to all other nodes in the network. Then, it is easy to show that

$$R_i = n\beta k\left[(\Delta x_i)^2 + \overline{(\Delta x)^2}\right] = nIC(i)^{-1}. \tag{32}$$

That is, the information centrality measure of a node is inversely proportional to the sum of the resistance distances from the corresponding node to all other nodes in the network.

## 7. Application to a social network

The first conclusion from the previous analysis is that for a given network the node displacement and the information centrality contain exactly the same topological information about the corresponding node, i.e., they are linearly related as

$$1/IC(i) = \beta k\left[(\Delta x_i)^2 + \overline{(\Delta x)^2}\right]. \tag{33}$$

Despite the relationship between the information centrality and the node displacement, there is a fundamental difference between them. The information centrality can be seen as a composite index containing local information about a node as well as global topological information about the network, whereas with the node displacement we can separate it into the local information as $(\Delta x_i)^2$ and the global one as $\overline{(\Delta x)^2}$. This difference is very relevant when comparing nodes in different networks.

Let us first see the difference using the simple example in Fig. 2. We listed the breakdown of $1/IC(i)$ into $(\Delta x_i)^2$ and $\overline{(\Delta x)^2}$. In terms of the information centrality, the vulnerability of the nodes 1 and 5 differs only by factor 1.6. In terms of the square node displacement, however, they differ by factor ~3. Moreover, we can claim the following. For the node 1, we see $(\Delta x_i)^2 > \overline{(\Delta x)^2}$, which suggests that the vulnerability of the node 1 comes mainly from its own weak position. For the node 5, on the other hand, we see $(\Delta x_i)^2 < \overline{(\Delta x)^2}$, which suggests that the vulnerability of the node 5 comes mainly from the weakness of the entire network, not from its own weakness. This observation suggests that the node displacement has a better resolution of the vulnerability than the inverse information centrality.

**Insert Tab. 1 about here.**

For a more realistic illustration we are going to use the data obtained by Kapferer about the social ties among tailor shops in Zambia [26]. Hereafter we use the parameter value of $\beta k = 1$, for which the vibrational energy is comparable to the thermal energy. The network is formed by 39 individuals who were observed during a period of ten months for their friendship and socio-emotional relationships. The data consists of two networks of social interactions recorded at two different times [26, 27]. After the first data was collected an abortive strike was reported [26]. Then, after seven months the second data was collected and a successful strike took place after it. In Fig. 4 we illustrate the social interactions between the 39 people in the tailor shop at the two different times.

**Insert Fig. 4 about here.**

We have proved analytically that the information centrality and the node displacement are directly related to each other for the nodes of a network. Then, we focus here on the differences between the two indices when we compare the same nodes in two different versions of the same network. For the purpose, we are going to use the change in the information centrality after seven months in the tailor shop $\Delta IC(i) = IC_2(i) - IC_1(i)$, where $IC_t(i)$ is the information centrality of node $i$ at time step $t$. In a similar way we define $\Delta\Delta x_i$ as the difference in the node displacement at the two times in which the network was analysed. In Fig. 5 we represent the differences in $\Delta IC(i)$ and $\Delta\Delta x_i$ for all nodes in the tailor shop. For the sake of comparison, the values of $\overline{(\Delta x)^2}$ for these two versions of the tailor shop network are 0.242 and 0.129, respectively.

**Insert Fig. 5 about here.**

This example clearly illustrates the differences between the information centrality and the node displacement in characterization of nodes. According to $\Delta IC(i)$ the individuals that displayed the largest change in their centrality are: Meshak > Adrian > Zakeyo > Chipalo >

Chipata > Kamwefu > Ibrahim > Mukubwa > Nkoloya > Enoch. However, the ranking according to the node displacements is: Zakeyo > Chipalo > Adrian > Sign > Enoch > Donal > Meshak > Seans > Kamwefu > Chipata. Meshak, who is ranked as number one according to $\Delta IC(i)$, is the actor displaying the largest change in the degree among all the actors. It changes its degree from 4 to 18 in the two versions of the tailor shop. However, Zakeyo and Chipalo, who are ranked by $\Delta\Delta x_i$ as the ones having the largest change in the node displacement have changed their degree from 1 to 7 in the two versions of the tailor shop network.

According to Eq. (33), the increase in the information centrality of a node can mean either the decrease in the local node displacement or the decrease in the global (average) node displacement (or both). Comparing the two panels in Fig. 5, we realise that the increase of the information centrality is, for most nodes, due to the decrease of the average node displacement, not due to the local one. Many nodes had many links in the first place and were not very vulnerable. Their information centrality increased after seven month mainly because the number of links generally increased in the entire network and hence $\overline{(\Delta x)^2}$ decreased, whereas their own node displacement scarcely decreased, that is, their own local vulnerability scarcely changed. For some nodes such as Zakeyo and Chipalo, however, their own vulnerability improved greatly after seven months because their own degrees increased dramatically. This demonstrates clearly that the node displacement has a better resolution of the situation than the information centrality.

In general, the change in the information centrality is very much correlated with the change in the degree centrality $DC(i)$. For instance, the correlation coefficient between $\Delta IC(i)$ and $\Delta DC(i)$ is 0.87 for the nodes in the network analysed here. However, $\Delta\Delta x_i$ displays a poor correlation coefficient of only 0.59 with $\Delta DC(i)$. One example that displays very well these differences is provided by the ranking of Sign, who is ranked as the fourth largest change in $\Delta\Delta x_i$ among all actors in the tailor shop. He is ranked as number 36 out of 39 according to the change in the information centrality and between 29 and 32 according to the change of the degree centrality, which is only 1. Sign was in a very vulnerable position in the initial network as he was connected to only one actor. After seven months he displayed a less vulnerable position in the network as he appears connected to two other actors. In the cases of Zakeyo and Chipalo this change has been more dramatic as they increased their degree from 1 to 7 in the two versions of the network and they appear as the ones having the largest changes in $\Delta\Delta x_i$.

## 8. Summary

In the present paper we proposed the node displacement as a measure of vulnerability of each node in a network. We defined the node displacement as the amplitude of vibration caused by thermal fluctuation of a heat bath. This simulates the situation where the network in question is under a level of external stress. We proved exact relations among the node displacement, the information centrality and the Kirchhoff index. The relation between the first two suggested that the node displacement has a better resolution of vulnerability than the information centrality; the latter is the sum of the local node displacement and the averaged one. We demonstrated this using the data of a social network of barbers in Zambia.

## References

[1] L.C. Freeman, Sociometry 40 (1977) 35.

[2] S.P. Borgatti, Social Networks 27 (2005) 55.

[3] S. Wasserman, K. Faust, Social Network Analysis, Cambridge University Press, Cambridge, 1994.

[4] H. Jeong, S.P. Mason, A.-L. Barabási, Z.N. Oltvai, Nature 411 (2001) 41.

[5] E. Estrada, Proteomics 6 (2006) 35.

[6] G. del Rio, D. Koschutzki, G. Coello, BMC Syst. Biol. 3 (2009) 102.

[7] F. Jordan, Phil. Trans. Royal Soc. B, Biol. Sci. 364 (2009) 1733.

[8] E. Estrada, Ö. Bodin, Ecol. Appl. 18 (2008) 1810.

[9] M. Barthelemy, A. Barrat, A. Vespignani, Adv. Compl. Syst. 10 (2007) 5.

[10] J.H. Choi, G.A. Barnett, B.S. Chou, Global Networks 6 (2006) 81.

[11] L.C. Freeman, J. Soc. Struct. 9 (2008) 1.

[12] P. Holme, B.J. Kin, C.N. Yoon, S.K. Hau, Phys. Rev. E 65 (2002) 056109.

[13] L. Zhao, K. Park, Y.-C. Lai, Phys. Rev. E 70 (2004) 035101.

[14] J. Wang, L. Dong, D.J. Hill, G.H. Zhang, Physica A 387 (2008) 6671.

[15] A. Santiago, R.M. Benito, Physica A 388 (2009) 2234.

[16] G. Chen, Z.Y. Dong, D.J. Hill, G.H. Zhang, Physica A 388 (2009) 4259.

[17] R. Albert, H. Jeong, A.-L. Barabási, Nature 406 (200) 378.

[18] K. Stephenson, M. Zelen, Social Networks 11 (1989) 1.

[19] M. Fiedler, Czech. Math. J., 23 (1973) 298.

[20] D.A. Spielman, S.-H. Teng, Lin. Algebra Appl. 421 (2007) 284.

[21] D.J. Klein, M. Randić, J. Math. Chem. 12 (1993) 81.

[22] W. Xiao, I. Gutman, Theor. Chem. Acc. 110 (2003) 284.

**[23]** Y.J. Yang, H.P. Zhang, J. Phys. A.: Math. Theor., 41 (2008) 445203.

[24] H.Y. Chen, F.J. Zhang, Disc. Appl. Math. 155 (2007) 654.

[25] B. Zhou, N. Trinajstić, J. Math. Chem. 46 (2009) 283.

[26] B. Kapferer, Strategy and transaction in an African factory. Manchester University Press, Manchester, 1972.

[27] http://vlado.fmf.uni-lj.si/pub/networks/data/Ucinet/UciData.htm.

**Table 1**

Tab. 1: Inverse of the information centrality and the squared node displacement for the nodes of the network shown in Fig. 2. These two measures satisfy Eq. (33) with $\overline{(\Delta x)^2} = 1.086$ and $\beta k = 1$.

| Nodes | Degree | $1/IC(i)$ | $(\Delta x_i)^2$ |
|---|---|---|---|
| 1, 2, 3, 7, 8, 9 | 1 | 2.445 | 1.357 |
| 4, 6 | 4 | 1.667 | 0.581 |
| 5 | 2 | 1.555 | 0.469 |

## Figure Captions

Fig. 1: A network submerged in a thermal bath. A “gas” at $T \to \infty$, a “liquid” at $T = 1$, and a “solid” at $T = 0$.

Fig. 2: A simple network for illustration.

Fig. 3: Correlation between the node degree and the node displacement. The solid curve indicates the power-law relation $\Delta x_i \sim k_i^{-1.155}$ , with correlation coefficient 0.958.

Fig. 4: The social network of 39 tailor shops in Zambia at one time (top) and seven months after that (bottom).

Fig. 5: Temporal changes in the information centrality and the node displacement. Each node is drawn with a diameter proportional to the absolute value of $\Delta IC(i)$ (top) or $\Delta \Delta x_i$ (bottom), respectively. In the upper panel, the red circles indicate the nodes where their information centrality increased (less vulnerability) and the blue ones indicate the nodes where it decreased (more vulnerability). In the lower panel, the red circles indicate the nodes where the node displacement increased (less vulnerability) and the blue ones indicate the nodes where it increased (more vulnerability).

**Fig. 1**

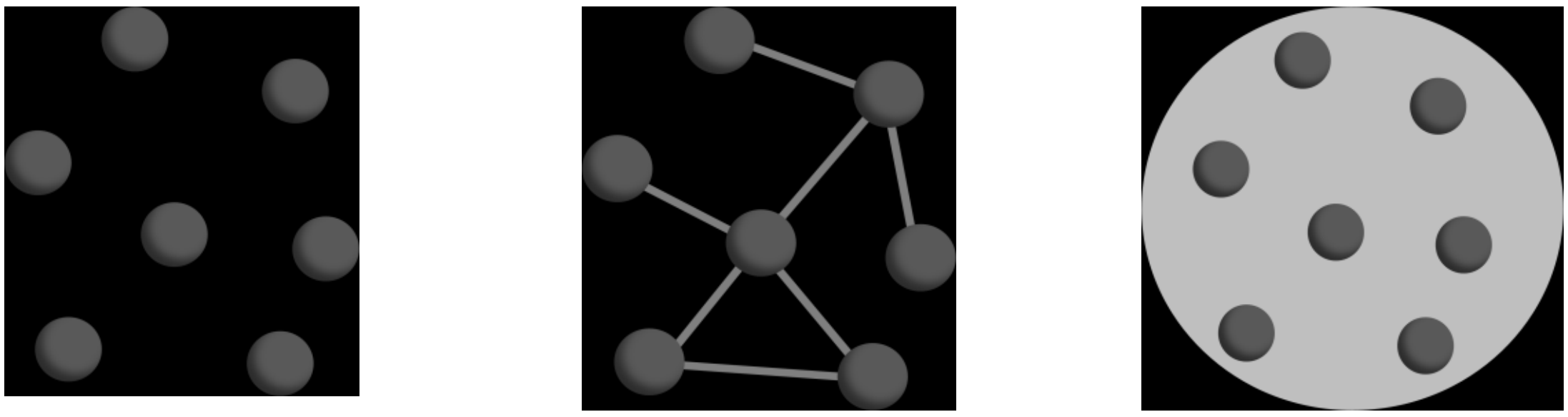

$T \to \infty$
$\beta \to 0$

$\beta = 1$

$T \to 0$
$\beta \to \infty$

Fig. 2

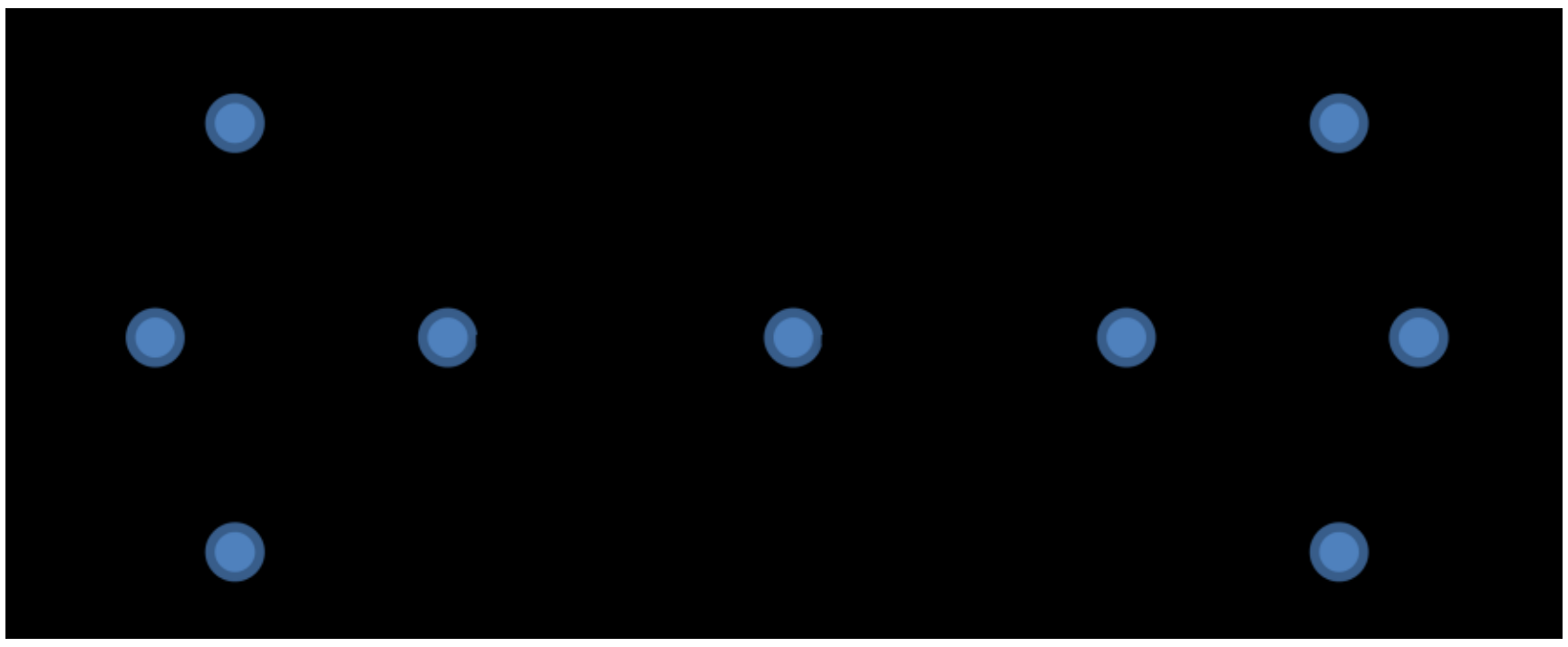

Fig. 3

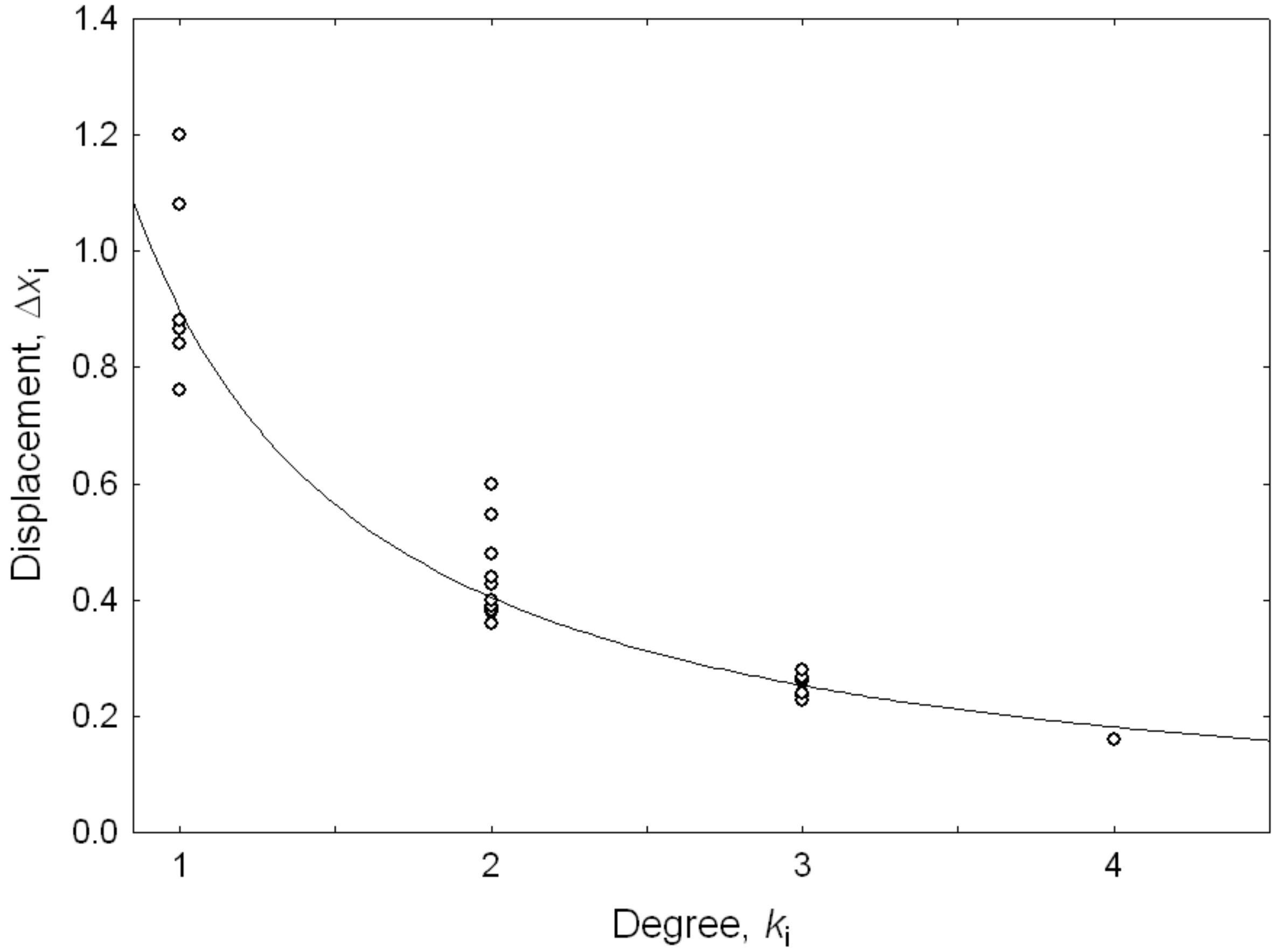

Fig. 4

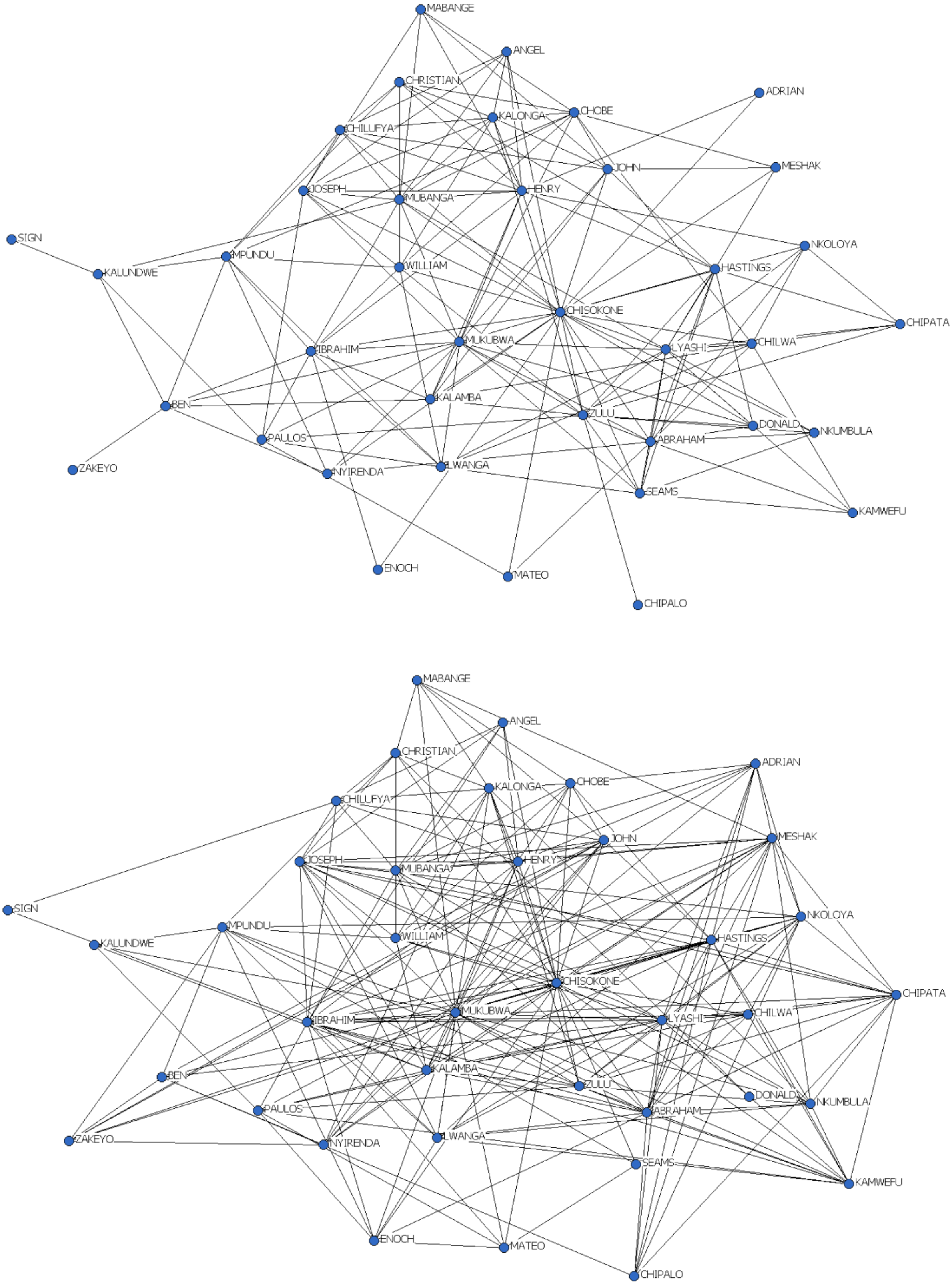
MABANGE
ANGEL
CHRISTIAN
ADRIAN
KALONGA
CHOBE
CHILUFYA
JOHN
MESHAK
JOSEPH
MUBANGA
HENRY
SIGN
NKOLOYA
MPUNDU
KALUNDWE
WILLIAM
HASTINGS
CHISOKONE
CHIPATA
IBRAHIM
MUKUBWA
LYASHI
CHILWA
KALAMBA
BEN
ZULU
DONALD
NKUMBULA
PAULOS
ABRAHAM
ZAKEYO
NYIRENDA
LWANGA
SEAMS
KAMWEFU
ENOCH
MATEO
CHIPALO
MABANGE
ANGEL
CHRISTIAN
ADRIAN
KALONGA
CHOBE
CHILUFYA
MESHAK
JOHN
JOSEPH
MUBANGA
HENRY
NKOLOYA
SIGN
MPUNDU
HASTINGS
KALUNDWE
WILLIAM
CHISOKONE
CHIPATA
MUKUBWA
CHILWA
IBRAHIM
LYASHI
KALAMBA
BEN
ZULU
DONALD
NKUMBULA
PAULOS
ABRAHAM
LWANGA
ZAKEYO
NYIRENDA
SEAMS
KAMWEFU
ENOCH
MATEO
CHIPALO

Fig. 5

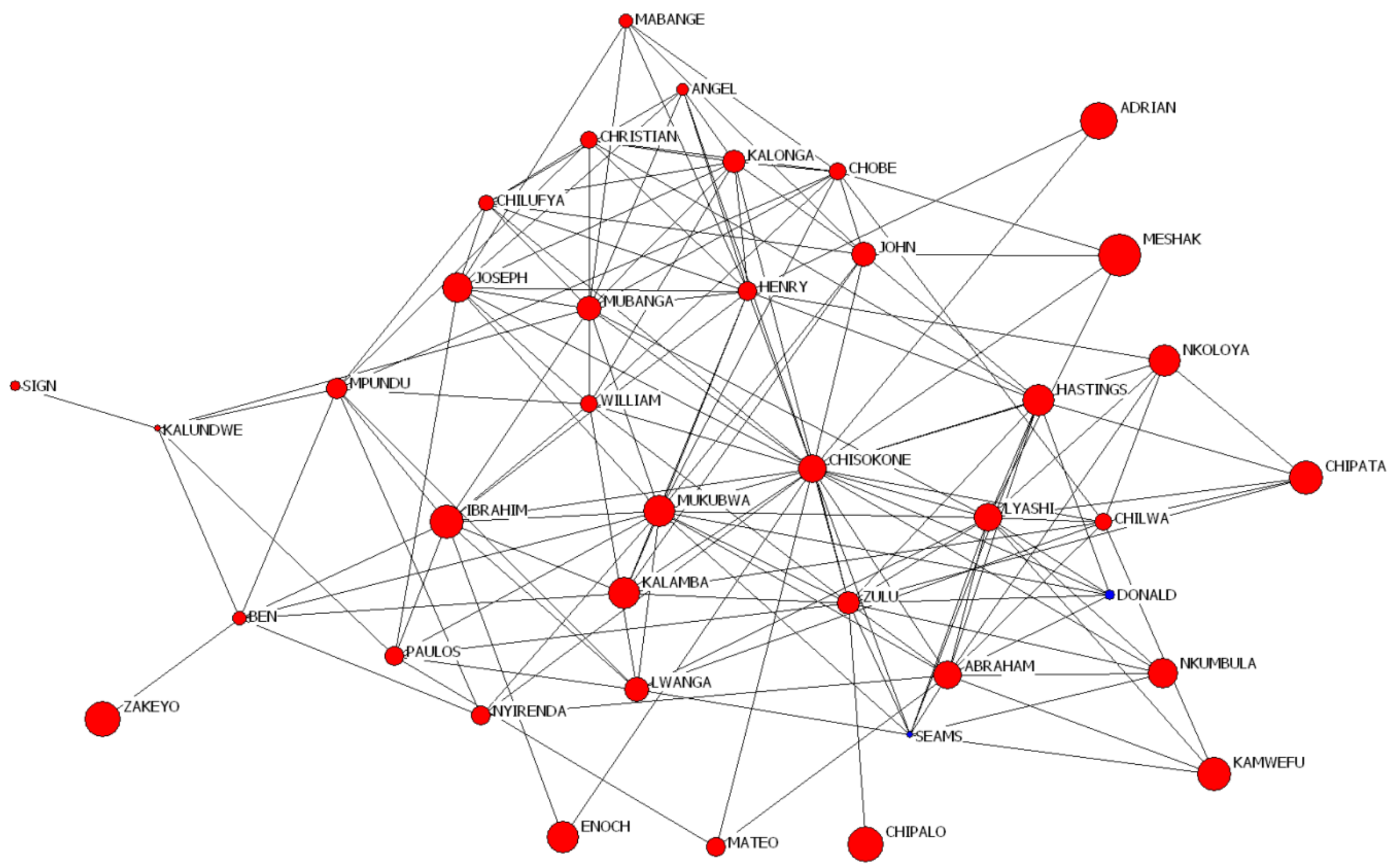

MABANGE
ANGEL
ADRIAN
CHRISTIAN
KALONGA
CHOBE
CHILUFYA
JOHN
MESHAK
JOSEPH
MUBANGA
HENRY
NKOLOYA
SIGN
MPUNDU
HASTINGS
KALUNDWE
WILLIAM
CHISOKONE
CHIPATA
IBRAHIM
MUKUBWA
LYASHI
CHILWA
KALAMBA
DONALD
ZULU
BEN
PAULOS
ABRAHAM
NKUMBULA
LWANGA
ZAKEYO
NYIRENDA
SEAMS
KAMWEFU
ENOCH
MATEO
CHIPALO

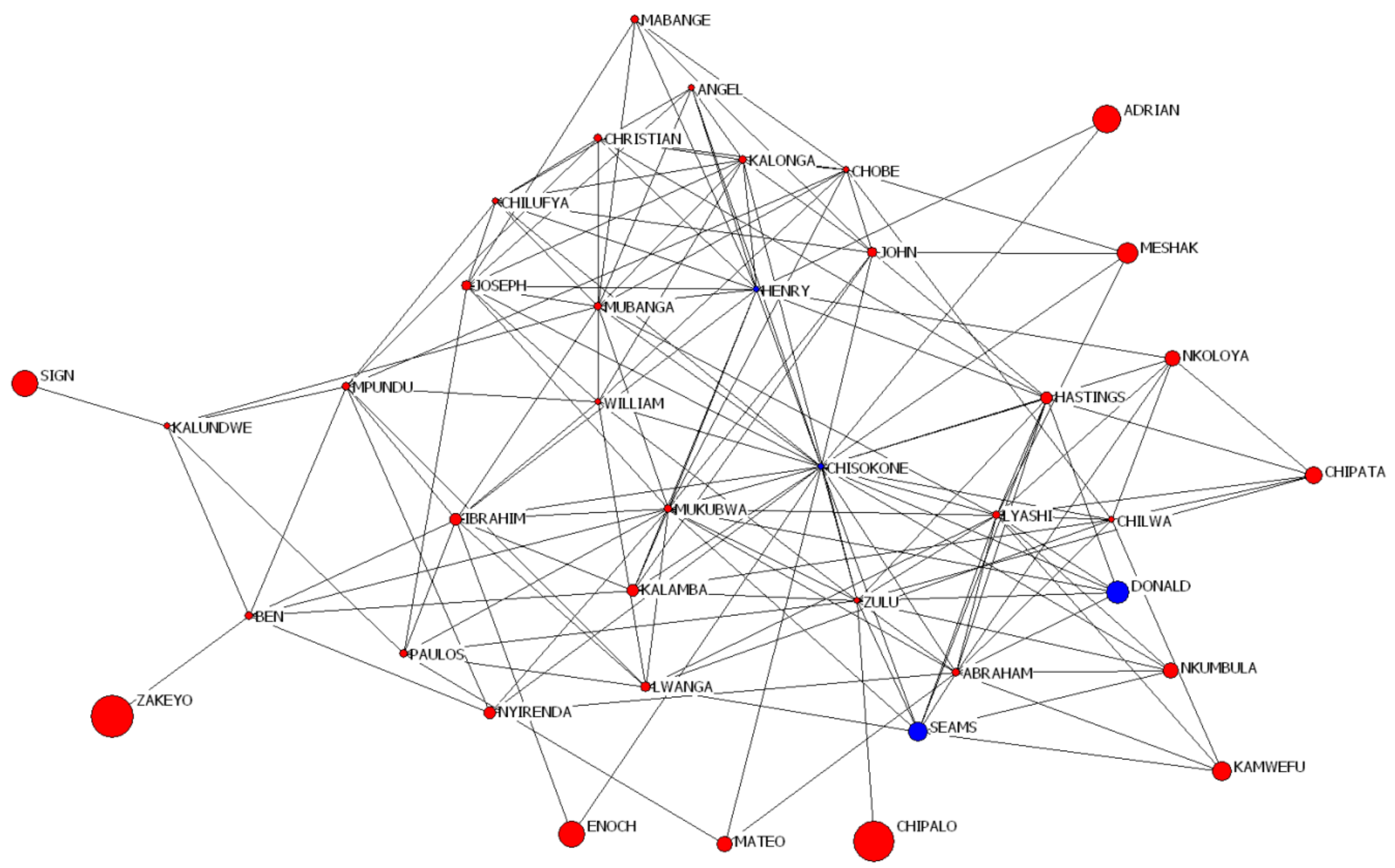

MABANGE
ANGEL
ADRIAN
CHRISTIAN
KALONGA
CHOBE
CHILUFYA
JOHN
MESHAK
JOSEPH
HENRY
MUBANGA
NKOLOYA
SIGN
MPUNDU
HASTINGS
KALUNDWE
WILLIAM
CHISOKONE
CHIPATA
LYASHI
IBRAHIM
MUKUBWA
CHILWA
KALAMBA
DONALD
ZULU
BEN
PAULOS
NKUMBULA
ABRAHAM
LWANGA
ZAKEYO
NYIRENDA
SEAMS
KAMWEFU
ENOCH
MATEO
CHIPALO